\documentclass{cimento}
\usepackage{graphicx}        
\usepackage{color}           
\usepackage{url}             

\title{Differential photometry of delta Scorpii during 2011 periastron}

\author{C.~Sigismondi\from{ins:s}\from{ins:i}\from{ins:a}\from{ins:u}\from{ins:e}}

   \instlist{\inst{ins:s} Sapienza University of Rome, P.le Aldo Moro 5 00185, Roma (Italy) 
              
             \inst{ins:i} ICRA, International Center for Relativistic Astrophysics, P.le Aldo Moro 5 00185, Roma (Italy)
                    
             \inst{ins:a} Universit\'e de Nice-Sophia Antipolis (France)
                     
             \inst{ins:u} Istituto Ricerche Solari di Locarno (Switzerland)
                    
             \inst{ins:e} GPA, Observatorio Nacional, Rio  de Janeiro (Brasil)                   
             }
\PACSes{\PACSit{95.75.Tv}{Digital imaging astronomy}
\PACSit{97.80.-d}{Binary stars}
\PACSit{97.30.-b}{Variable stars}
\PACSit{26.30.Ca}{Accreting binary systems explosive burning}}

\begin{document}

\maketitle

\begin{abstract}
Hundred observations of Delta Scorpii over 200 days, from April 2 to October 16, 2011, have been made for AAVSO visually and digitally from Rio de Janeiro, Rome and Paris. The three most luminous pixels either of the target star and the two reference stars are used to evaluate the magnitude through differential photometry. The main sources of errors are outlined. The system of Delta Scorpii, a spectroscopic double star, has experienced a close periastron in July 2011 within the outer atmospheres of the two giant components. The whole luminosity of Delta Scorpii system increased from about Mv=1.8 to 1.65 peaking around 5 to 15 July 2011, but there are significant rapid  fluctuations of 0.2 - 0.3 magnitudes occurring in 20 days that seem to be real, rather than a consequence of systematic errors due to the changes of reference stars and observing conditions. This method is promising for being applied to other bright variable stars like Betelgeuse and Antares. After August the magnitude remained constant at Mv=1.8 until the last observation on October 16 made in twilight from Rome.
\end{abstract}

\section{Introduction}

Delta Scorpii is a double giant Be star in the forefront of the Scorpio, well visible to the naked eye, being normally of magnitude 2.3. In the year 2000 its luminosity rose up suddenly to the magnitude 1.6, changing the usual aspect of the constellation of Scorpio. This phenomenon has been associated to the close periastron of the companion, orbiting on an elongate ellipse with a period of about 11 years. The new periastron, on basis of high precision astrometry, occurred in the first decade of July 2011, with the second star of the system approaching the atmosphere of the primary, whose circumstellar disk has a H-alpha diameter of 5 milliarcsec, comparable with the periastron distance \cite{Meilland}. The results of a photometric campaign visual and digital, with observations made in Rio de Janeiro and in Rome are presented. The method of data analysis of the differential photometry, based on Poisson statistics, it is suitable to be used also with other bright variable stars like Betelgeuse by a large public and with commercial cameras.  This is in the spirit of fostering the "citizen astronomy" of AAVSO. 

\section{Why Delta Scorpii?} 
When in Rio de Janeiro, visiting the Observatorio Nacional, I realized that in my home at 142 rua Marqu\^es de Abrantes, the window aimed exactly the azimut of the celestial objects rising to the zenith: Delta Scorpii shined there during one clear evening of March 2011 and I identified it from this remarkable property. Only later with the apparition of Antares I recognized the usual geometry of Scorpio, upside down because of the Southern hemisphere. 

The forthcoming periastron of delta Sco expected in July 2011 \cite{Meilland,Sigismondi1107,Ames} led me to monitor this star, and the AAVSO to select it as the star of that month \cite{Otero}.

\medskip
Delta Scorpii is an easy target to be found with naked eyes, it is visible even in urban contexts with heavy luminous pollution, and therefore it opens the field of naked eye variable stars which is far from being saturated by the "citizen astronomers".

\section{Differential photometry for visual observations and airmass corrections}

The choice of beta ($Mv=2.50$, from Simbad website\footnote[1]{\url{http://simbad.u-strasbg.fr/simbad/}}) and pi ($Mv=2.89$) Scorpii as reference stars for the images taken with my CMOS is due to their color index $B-V\sim 0$, similar to the one of delta Scorpii. When pi Scorpii was too low and not visible because of the hazes near the horizon I used alfa Scorpii ($Mv=1.09$, but also slightly variable) as reference.

\medskip
When, for visual observations, I changed the reference star to eta Ophiuchi ($Mv=2.43$), changes occurred in the estimated magnitude. Alpha Ophiuchi ($Mv=2.10$) was a better reference \cite{Otero}, even if farther than eta Oph.

\medskip
With the star near the horizon there are great differential effects even for small angular distances like between beta and delta Scorpii. 

\medskip
Near the horizon the airmass estimate with Garstang's \cite{Garstang} model does not fully explain the loss of luminosity of the star with respect to its references: when for visual observations I used alpha Ophiuchui,  to apply the airmass correction I preferred the classical cosecant law, even with its infinity at zero altitude: $\Delta M=0.13/[sin(h_{1})- sin(h_{2})]$ with  $h_{1}$ the height of delta Sco and $h_{2}$ of alpha Oph.  A layer of humid air near the ground after sunset in summer time, for observations made in the last few degrees above the horizon can act as 60 airmasses of clear atmosphere \cite{Sigismondi1106} while for Garstang there are only 5 airmasses which determine a loss of $\Delta M=0.65$ magnitudes. 

\medskip
Moreover I have selected reference pairs of stars in order to compare visually with them the difference in magnitude between delta, alpha and beta Scorpii. These pairs are alpha and beta Centauri ($\Delta Mv=0.70$)  for the Southern hemisphere and alpha Lyrae and alpha Aquilae ($\Delta Mv=0.74$) for the Northern one. Delta was for most of the time exactly intermediate  (i.e. of $Mv=1.8$) between alpha and beta so that the difference between alpha and delta was $\Delta Mv=0.7$ and the difference between delta and beta was also $\Delta Mv=0.7$.

\medskip
Scatters as large as 0.4 magnitudes, in the evaluation of the magnitude of delta, could be attributed to the humid layer and to the skyglow.  The observations have been done all in the lights of big cities like Rio de Janeiro, Rome, Pescara, Nice and Paris: moreover in the last three cities the observations were made at less than $25^\circ$ above the horizon.

Delta Scorpii undergoes variations of luminosity with a 60 days period \cite{Otero}, this could explain the changes in luminosity of $\Delta M = 0.2 - 0.3$ magnitudes observed within 20 days. 	

There are visual data where the magnitude of delta Sco drops down to $Mv = 2.2$ in the end of July 2011. I have attributed this phenomenon to the change of reference star, but also this drop and the following rising can be real. There are also digital data ($Mv=2.27$ on June 14.05, 2.17 on July 11.9 and 2.18 on August 7.8 all with the Moon at $1^\circ - 2^\circ$, even if occulted) which seem to be spurious.

\begin{figure}
\centerline{\includegraphics[width=1\textwidth,clip=]{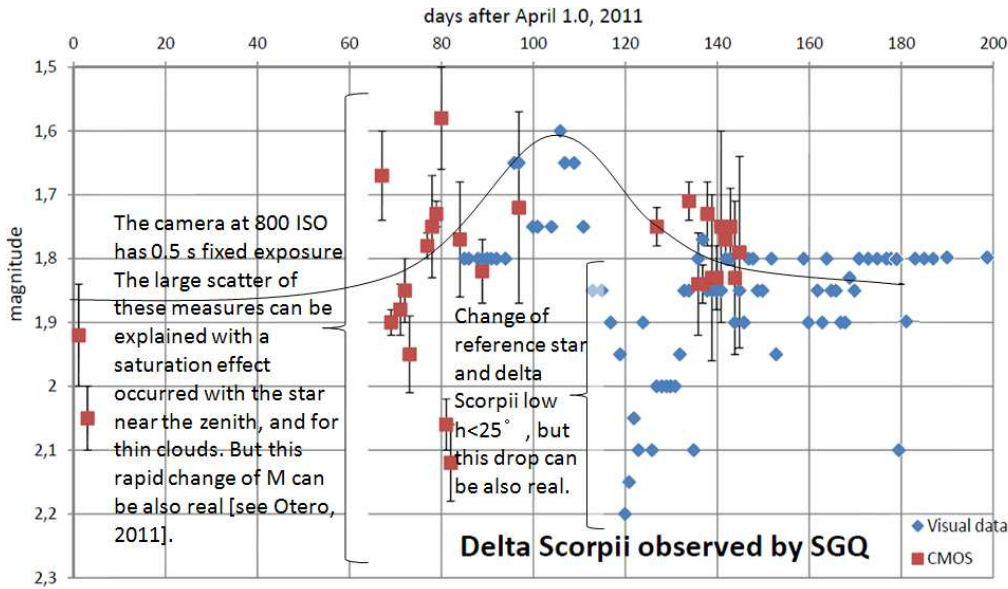}}
\caption{Plot of hundred observations of Delta Scorpii: visual (blue diamonds) and digital (red squares). The star reaches its maximum luminosity $Mv \sim 1.65$ around July 5-15, 2011. The rest of the time the luminosity remains around the magnitude $Mv \sim 1.8$, with some fluctuations, possibly physical, of $\Delta M=0.2 - 0.3$ magnitudes . The digital photo were made with the portable camera SANYO CG9 CMOS detector at 800 ISO, 0.5 s of exposure and 3x of visual zoom. The regime of linearity of such detector has been verified within 0.1 magnitudes using a photo of the Southern Cross, made with the same exposure, ISO and magnification.}
\label{Fig. 1}
\end{figure}

\begin{figure}
\centerline{\includegraphics[width=1\textwidth,clip=]{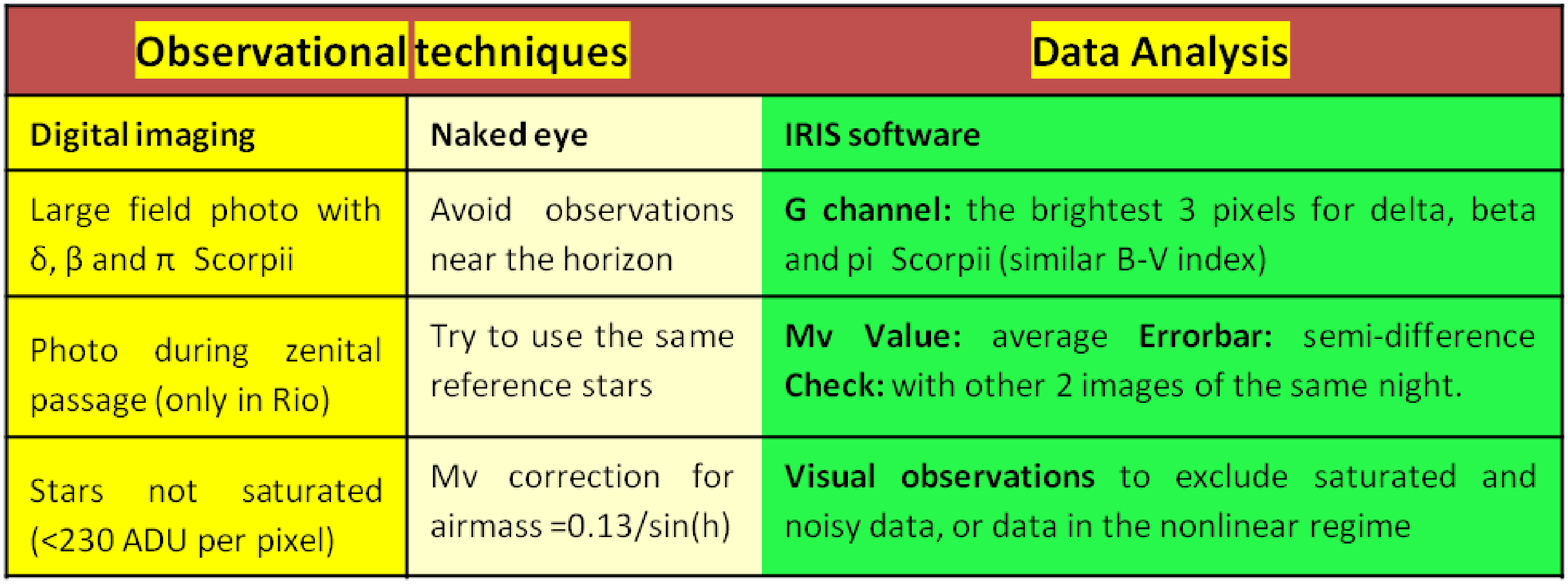}}
\caption{The scheme of the techniques adopted for the observations and for the analysis of them.}
\label{Fig. 2}
\end{figure}

\begin{figure}
\centerline{\includegraphics[width=0.7\textwidth,clip=]{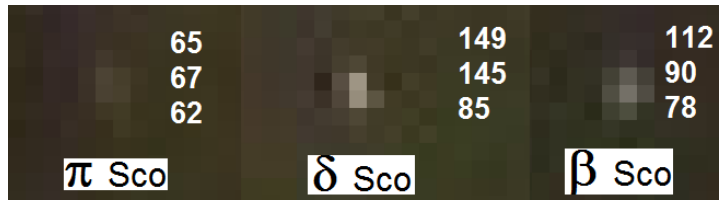}} 
\caption{The stars with their brightest 3 pixels, used for the estimation of the magnitude. Here is the image of July 12, 2011, with Delta very close to the full Moon. The intensity of each star is proportional to the sum $\Sigma$ of the three brightest pixels in the G channel. The relative error of such hypothesis of proportionality to the sum $\Sigma$, due to Poisson's statistics, is $1/\sqrt{\Sigma}$. This channel is the closest to Johnson's V band \protect\cite{Lanciano}.  The luminosity of delta Scorpii with respect to its reference stars $\beta$ and $\pi$  is $Mv\beta,\pi-2.5\log{(\Sigma\delta/\Sigma\beta,\pi)}$. The value published in the AAVSO website is the average $(Mv\beta+Mv\pi)/2$ and the errorbar adopted is the semi-difference between $Mv\beta$ and  $Mv\pi$.}
\label{Fig. 3}
\end{figure}

\section{Conclusions}

The digital method requires always calibrations and the stars have not to saturate the detectors. Moreover the detectors have to work in their linear regime.
  
\medskip
The visual method shows that the errors are reduced by using always the same reference stars. 

\medskip
Beyond some real physical phenomenon appearing, the intrinsic scatter of the databases of such stars shows the complexity of the task to reduce the visual and digital errorbar below 0.1 magnitudes. 

\medskip
Alpha Orionis and Alpha Scorpii itself, are very interesting targets of similar observational campaigns, as well as other bright stars. Being the brightest stars in their surroundings the calibration of the detectors is crucial for the success of these studies.


\bibliography{DeltaScorpii}
   
\bibliographystyle{varenna}

\end{document}